\documentclass[%
reprint,
amsmath,amssymb,
aps,
prb,
superscriptaddress
]{revtex4-2}
\usepackage[utf8]{inputenc}
\usepackage{multirow}
\usepackage{braket}
\usepackage{mathtools}
\usepackage{babel}
\usepackage{graphicx}
\usepackage{dcolumn}
\usepackage{bm}
\usepackage{xcolor}
\usepackage{xspace}
\usepackage{hyperref}
\usepackage[capitalize]{cleveref}
\usepackage[normalem]{ulem}

\newcommand{\scgw}{sc$GW$\xspace}
\newcommand{\gw}{\ensuremath{GW}\xspace}
\newlength{\templen}

\begin{document}
\title{Challenges with relativistic $GW$ calculations in solids and molecules}  
\author{Gaurav Harsha}
\author{Vibin Abraham}%
\affiliation{%
Department of Chemistry, University of Michigan, Ann Arbor, Michigan 48109, USA
}%
\author{Dominika Zgid}%
\affiliation{%
Department of Chemistry, University of Michigan, Ann Arbor, Michigan 48109, USA
}%
\affiliation{%
Department of Physics, University of Michigan, Ann Arbor, Michigan 48109, USA
}%

\date{\today}

\begin{abstract}
For molecules and solids containing heavy elements, accurate electronic structure calculations require accounting not only  for electronic correlations but also for relativistic effects. In molecules, relativity can lead to severe changes in the ground-state description. In solids, the interplay between both correlation and relativity can change the stability of phases or it can lead to an emergence of completely new phases.
Traditionally, the simplest illustration of relativistic effects can be done either by including pseudopotentials in non-relativistic calculations or alternatively by employing large all-electron basis sets in relativistic methods.
By analyzing different electronic properties (band structure, equilibrium lattice constant and bulk modulus) in semiconductors and insulators, we show that capturing the interplay of relativity and electron correlation can be rather challenging in Green's function methods.
For molecular problems with heavy elements, we also observe that  similar problems persist.
We trace these challenges to three major problems: deficiencies in pseudopotential treatment as applied to Green's function methods, the scarcity of accurate and compact all-electron basis sets that can be converged with respect to the basis set size, and linear dependencies arising in all-electron basis sets particularly when employing Gaussian orbitals.
Our analysis provides detailed insight into these problems and opens a discussion about potential approaches to mitigate them.
\end{abstract}

\maketitle

\section{Introduction}
Computational methods in \textit{ab initio} electronic structure theory have become indispensable tools in the design and study of new functional materials and molecules.
For many years, mean-field methods such as density functional theory~\cite{kohn_ground-state_1960,kohn_self-consistent_1965} (DFT) and Hartree-Fock (HF) were the commonly employed simulation tools mostly due to their low computational cost and reasonable accuracy.
However, recently, due to advances in computational resources as well as algorithms, sophisticated methods capturing electronic correlation beyond the mean-field approximation have gathered much more attention. 
Examples include coupled cluster,~\cite{crawford_introduction_2000,bartlett_coupled-cluster_2007} perturbation theories for wave function and Green's function,~\cite{hedin_new_1965,dahlen_self-consistent_2005,phillips_communication_2014,rusakov_self-consistent_2016,chen_random-phase_2017} and embedding methods~\cite{georges_dynamical_1996,lichtenstein_finite-temperature_2001,RevModPhys_lda_dmft,knizia_density_2012,kananenka_systematically_2015,lan_communication_2015,wouters_practical_2016,zgid_finite_2017,rusakov_self-energy_2019,sun_finite-temperature_2020} among others.

The development of new electronic structure methods is primarily driven by one or both of the following objectives: (i) accurate description of electron correlation, (ii) applicability to realistic systems with large number of electrons.
Addressing both of these challenges satisfactorily is usually very difficult. 
For instance, low-scaling mean-field methods such as DFT and HF (with $\mathcal{O}(N^3)$ and $\mathcal{O}(N^4)$ computational costs, respectively, $N$ being a measure of system size) can be applied to very large systems.
However, they often lack quantitative accuracy.
On the other hand, coupled cluster with single, double and perturbative triple excitation, i.e., CCSD(T), is considered as the gold standard for description of dynamical correlation, but has limited applicability due to its high $\mathcal{O}(N^7)$ scaling.
Consequently, greater recovery of correlation effects usually comes hand-in-hand with an increased computational cost.

The \gw method,~\cite{hedin_new_1965,aryasetiawan_thegwmethod_1998,golze_gw_2019,reining_gw_2018} derived from Hedin's perturbation theory for single-particle Green's function, offers a reasonable balance between accuracy and cost.
In comparison to DFT or HF, it provides a high-level description of correlation effects at a formal scaling of $\mathcal{O} (N^6)$, which is usually brought down to $\mathcal{O} (N^4)$ by using  approximations such as the density-fitting procedure.~\cite{vahtras_integral_1993,sun_gaussian_2017,ye_fast_2021}
The good accuracy of \gw can be explained by its diagrammatic similarities with theories like coupled cluster and random phase approximation.~\cite{scuseria_particle-particle_2013,lange_relation_2018,quintero-monsebaiz_connections_2022,tolle_exact_2023}
Additionally, \gw has shown versatility in its application to both finite~\cite{stan_fully_2006,caruso_unified_2012,marom_benchmark_2012,van_setten_gw-method_2013,van_setten_gw100_2015,caruso_benchmark_2016,forster_low-order_2021,forster_two-component_2023,wen_comparing_2023,abraham_relativistic_2024} and extended systems.~\cite{faleev_all-electron_2004,shishkin_self-consistent_2007,Kutepov2009,Kutepov2017,kutepov_ground_2017,grumet_beyond_2018}

As  the applications of computational methods are expanded to systems with heavy elements, another challenge arises: the problem of accurate inclusion of relativity.
Even for elements in third and fourth row of periodic table, relativistic effects are important.~\cite{pyykko_relativistic_2012}
While for a full relativistic description, the 4-component Dirac equation, instead of the regular non-relativistic Schr\"odinger equation, should be solved, solving the 4-component equation is a formidable task and not strictly necessary for most systems if only the electronic structure properties are of interest.
Several approximations have been proposed to introduce the full relativistic physics into an effective Hamiltonian that describes only the electronic degrees of freedom.
The most widely used approximation follows the work of Breit,~\cite{breit_diracs_1932} where the Dirac equation is reduced to Schr\"odinger-like equation with an effective Dirac-Coulomb-Breit Hamiltonian that contains the relativistic terms.
Theories such as the Douglas-Kroll-Hess~\cite{douglas_quantum_1974,hess_applicability_1985,hess_relativistic_1986,nakajima_douglaskrollhess_2012,reiher_douglaskrollhess_2014} and the exact two-component~\cite{kutzelnigg_quasirelativistic_2005,liu_quasirelativistic_2007,liu_exact_2009,sun_exact_2009,cheng_analytic_2011} (X2C) introduce further approximations to make relativistic calculations achievable.
Several \textit{ab-initio} implementations for realistic systems of the relativistic \gw theory have emerged in the condensed matter community,~\cite{sakuma_gw_2011,kutepov_electronic_2012,scherpelz_implementation_2016,holzer_ionized_2019} and in the quantum chemistry community,~\cite{kuhn_one-electron_2015,yeh_relativistic_2022,forster_two-component_2023} where the latter are based on the X2C Hamiltonian.

Calculations in the X2C approach are generally performed using all-electron basis sets since current pseudopotentials (PPs) or effective core potentials (ECPs) do not recover relativistic effects such as spin-orbit coupling (SOC).
Most modern implementations or ab-initio relativistic codes are capable of handling relativistic elements, at least up to 5th or 6th row, including the Lanthanides, in the periodic table.
Yet, performing reliable \gw calculations for accurate electronic structure properties still remains an arduous task.
Considering that the \gw approximation is the simplest in the hierarchy of Hedin's perturbation theory, it offers an understandably limited accuracy in describing electron correlation.
However, factors other than the accuracy of the GW method, e.g., poor quality and availability of basis sets, lack of effective PPs, etc., can have a larger impact on the quality of results, particularly in heavy elements.

In this discussion article, we analyze the factors that hinder the applicability of the sophisticated machinery of \gw to both solids and molecules containing heavy elements.
In particular, we first highlight the need for all-electron relativistic calculations due to the inability of PP to account for many important relativistic effects.
We then investigate problems with achieving the basis set convergence in $GW$ calculations.
This is followed by results that highlight arising linear dependencies for larger basis sets. 
Such problems often aggravate the problem of slow convergence with respect to basis set size.
Finally, we also comment on possible improvements of experimental results that are ultimately necessary for benchmarking and validating new theoretical methods.
By investigating multiple properties of solids and molecules, we perform a comprehensive analysis of how several external factors, other than the accuracy of the method itself, influence the results in self-consistent \gw (\scgw).

This article is organized as follows: in \cref{sec:methods}, we recap some of the theoretical concepts on which majority of this paper is based.
This is followed by computational and implementation details in \cref{sec:comp-details}.
In \cref{sec:results}, we analyze various hurdles, one by one, for the application of \scgw to relativistic systems.
A general discussion and conclusions on possible factors behind these challenges, along with the current and future developments that can address these problems, is presented in \cref{sec:discussion_conclusion}.

\section{\label{sec:methods}Methods}
In this work, we employ mean-field methods and self-consistent $GW$ (\scgw) theory within the X2C relativistic formalism.~\cite{kutzelnigg_quasirelativistic_2005,liu_quasirelativistic_2007,liu_exact_2009,sun_exact_2009,cheng_analytic_2011}
While most of the theoretical and implementation details can be found in Refs.~\cite{yeh_relativistic_2022,yeh_fully_2022,abraham_relativistic_2024}, here we provide a brief overview of important concepts, namely the relativistic Hamiltonian, \scgw theory, and Birch-Murnaghan~\cite{murnaghan_compressibility_1944,birch_finite_1947} equation of state that we use to study bulk properties in materials.
Note that while GW by itself~\cite{golze_gw_2019,vanschilfgaarde_quasiparticle_2006,faleev_all-electron_2004,kotani_quasiparticle_2007,cao_fully_2017,Faleev2004,Schilfgaarde2006,pham_gw_2013,Govini_GW_large_scale,yeh_fully_2022} can be used for treating weakly correlated compounds with excellent results, it is also important for embedding approaches~\cite{bierman_gw+dmft,Multi-tier-GW-DMFT,Choi2016,embedding_in_GW,Zhu_GW+DMFT,Iskakov_MnO_NiO,Yeh_SrMnO} where it is usually employed as a low level method for treating the environment.

\subsection{\label{subsec:relativity}Relativistic Hamiltonian: exact two-component formalism}
The X2C formalism is widely used among electronic structure relativistic methods.~\cite{liu_exact_2009,cheng_analytic_2011,Liu2018,sun_exact_2009}
In this formalism, the four-component one-body Dirac-Coulomb Hamiltonian $H^{4C1e}$ is used as a starting point on which a unitary transformation is performed that decouples the eigenvectors with large and small eigenvalues, generally labeled as large and small components,
\begin{align}
    U^{\dagger} {H}^{\text{4C1e}} U
    &= 
    U^{\dagger} \left[\begin{array}{cc}
        \hat{V} & \hat{T} \\
        \hat{T} & \frac{1}{4 c^2} \hat{L}-\hat{T},
    \end{array}\right] U,
    \nonumber
    \\
    &=  \left[\begin{array}{cc}
        {H}_{+}^{\text{X2C1e}} & 0 \\
        0 & {H}_{-}^{\text{X2C1e}}
    \end{array}\right].
    \label{eq:x2c-ham}
\end{align}
Here, $\hat{V}$ is the local Coulomb potential, $\hat{T}$ is the kinetic energy matrix, and the $\hat{L}$ operator is defined as
\begin{equation}
    \hat{L} = \left( \vec{\sigma} \cdot \vec{p} \right)
    \hat{V}
    \left( \vec{\sigma} \cdot \vec{p} \right).
    \label{eq:x2c-operator}
\end{equation}
The assumption invoked in the X2C formalism is that after the unitary transformation, the large-component can be effectively considered as electron, even though in principle, it mixes the electronic and positronic components.
Therefore, the large component Hamiltonian $H_{+}^{X2C1e}$ is used as the one-body Hamiltonian in electronic structure calculations, along with the usual two-electron integrals.
We should note that no relativistic correction is used for the latter.

The formalism described so far is called the X2C1e approximation.
By definition, this approximation breaks the $S_z$ spin-symmetry and necessitates the use of generalized spin-orbitals in subsequent calculations.
Another approximation, called as the spin-free X2C1e (sfX2C1e) is also popular.
In sfX2C1e, only the spin-independent contribution arising form the $\hat{L}$-operator in \cref{eq:x2c-operator} is considered.
Given that sfX2C1e leads to a symmetry-adapted Hamiltonian, for several systems, where negligible or no SOC is present, using sfX2C1e greatly reduces the computational cost and memory requirements in a calculation, without any accuracy loss.
Even when SOC cannot be neglected, total energy trends in sfX2C1e do not differ significantly from X2C1e.~\cite{abraham_relativistic_2024}

\subsection{\label{subsec:scgw}Self-consistent $GW$}

Hedin's equations~\cite{hedin_new_1965} define a perturbation theory where the one-particle Green's function $G$, the vertex function $\Gamma$, the irreducible polarizability $\Pi_0$, the screened Coulomb interaction $W$, and the self-energy $\Sigma$ are related through a set of integro-differential equations.
This is traditionally represented as Hedin's pentagram diagram.
In the $GW$ approximation, higher-order corrections to the vertex arising from $G$ are ignored and the vertex $\Gamma$ is approximated as a Dirac delta function in space-time.
As a result, the self-energy is expressed as a product of the interacting Green's function $G$ and the screened Coulomb potential $W$.

Hedin's equations or the $GW$ approximation can be evaluated either on the real or imaginary axis. 
Here, we work with the finite-temperature (or Matsubara imaginary axis) formalism described in Refs.~\onlinecite{yeh_relativistic_2022,yeh_fully_2022}.
The one-particle imaginary-time Green's function is defined as
\begin{equation}
    G_{p \sigma, q \sigma'}^{\mathbf{k}} (\tau) = \frac{1}{\mathcal{Z}} \mathrm{Tr}
    \left(
        e^{- (\beta - \tau) (H - \mu N)} c_{p \sigma}^{\mathbf{k}} c_{q\sigma'}^{\mathbf{k},\dagger}
    \right),
    \label{eq:imag-time-gf}
\end{equation}
where $\beta = 1 / k_B T$ is the inverse temperature ($k_B$ being the Boltzmann constant), $\mu$ is the chemical potential, $H$ and $N$ are the Hamiltonian and particle number operators, $c_{p\sigma}^{\mathbf{k}}$ and $c_{p\sigma}^{\mathbf{k}, \dagger}$ are the second-quantization annihilation and creation operators for the $p$th Bloch orbital with the spin $\sigma$ and momentum label $\mathbf{k}$, and $\tau \in \left[0, \beta\right]$ is the imaginary time.
Finally, $\mathcal{Z}$ is the partition function, defined as
\begin{equation}
    \mathcal{Z} = \mathrm{Tr} \left(
        e^{-\beta (H - \mu N)}
    \right).
\end{equation}
In the $GW$ approximation, the self-energy on the imaginary time/frequency axis is approximated as
\begin{equation}
    \mathbf{\Sigma}^{\mathbf{k}} [\mathbf{G}] (i\omega_n)
    =
    \mathbf{\Sigma}^{\mathbf{k}}_\infty [\mathbf{G}]
    + \mathbf{\Sigma}^{\mathbf{k}}_c [\mathbf{G}] (i \omega_n),
    \label{eq:self-energy-v1}
\end{equation}
where $\mathbf{\Sigma}^{\mathbf{k}}_\infty$ is the static (frequency independent) Hartree-Fock self-energy, and $\mathbf{\Sigma}^{\mathbf{k}}_c (i\omega_n)$ is the dynamic contribution, defined in the imaginary time ($\tau$) axis as
\begin{multline}
    \left(\Sigma_c\right)^{\mathbf{k}}_{p\sigma, q\sigma'} (\tau)
    =
    \\
    -\frac{1}{N_{\mathbf{k}}} \sum_{\mathbf{q}} \sum_{ab}
    G^{\mathbf{k - q}}_{a\sigma, b\sigma'} (\tau)
    \tilde{W}_p^{\mathbf{k}} {}_a^{\mathbf{k-q}} {}_b^{\mathbf{k - q}} {}_q^{\mathbf{k}} (-\tau),
    \label{eq:self-energy-v2}
\end{multline}
where $\tilde{W}$ is the effective screened Coulomb interaction, $N_{\mathbf{k}}$ is the number of $k$-points sampled in the Brillouin zone (BZ).
The summations are performed over momentum vectors $\mathbf{q}$ in BZ, as well as the atomic orbitals $a$ and $b$ in the unit cell.
The new Green's function is then defined by the Dyson equation,
\begin{equation}
    \left[ \mathbf{G}^{\mathbf{k}} (i \omega) \right]^{-1}
    = (i\omega + \mu) \mathbf{S}^{\mathbf{k}} - \mathbf{H}_0^{\mathbf{k}} - \mathbf{\Sigma}^{\mathbf{k}} (i \omega),
    \label{eq:dyson}
\end{equation}
where the chemical potential $\mu$ is fixed to ensure a correct particle number, $\mathbf{S}^{\mathbf{k}}$ is the overlap matrix, and $\mathbf{H}_0^{\mathbf{k}}$ is the one-electron Hamiltonian.
We refer the readers to Ref.~\onlinecite{yeh_fully_2022} for exact expressions for the screened Coulomb operator $\tilde{W}$.
The total electronic energy is calculated using the Galitskii-Migdal formula,
\begin{subequations}
    \label{eq:total-energy}
    \begin{align}
        E &= E_{1b} + E_{2b},
        \\
        E_{1b} &= \frac{1}{2 N_{\mathbf{k}}} \sum_{\mathbf{k}}
        \mathrm{Tr} \left[
            \mathbf{\gamma}^\mathbf{k} \left(
                \mathbf{F}^{\mathbf{k}} + \mathbf{H}_{0}^{\mathbf{k}}
            \right)
        \right],
        \\
        E_{2b} &= \frac{2}{\beta N_{\mathbf{k}}} \sum_{n, \mathbf{k}}
        \mathrm{Tr} \left[
            \mathbf{G}^{\mathbf{k}} (i \omega_n) \mathbf{\Sigma}_c^{\mathbf{k}} (i \omega_n)
        \right]
    \end{align}
\end{subequations}
where $\mathbf{F}^{\mathbf{k}}$ is the Fock matrix, defined as
\begin{equation}
    \mathbf{F}^{\mathbf{k}} = \mathbf{H}_0 + \mathbf{\Sigma}_\infty^{\mathbf{k}},
    \label{eq:fock}
\end{equation}
and $\gamma^{\mathbf{k}} = -\mathbf{G}^{\mathbf{k}} (\tau = \beta^-) $ is the one-particle density matrix.
In our implementation of \scgw, \cref{eq:self-energy-v1,eq:self-energy-v2,eq:dyson} are solved iteratively until self-consistency is achieved in both the one-body contribution to the energy $E_{1b}$ as well as the total energy $E$.
Finite-size corrections for the static and dynamic contributions to the self-energy have been included, as described in Ref.~\onlinecite{yeh_fully_2022}.
Relativistic effects at the level of both spin-free and full X2C formalism with one-electron approximation (sfX2C1e and X2C1e) have been used.

\subsection{\label{subsec:eos}Equation of state for solids}
For periodic solids, the relationship between energy and volume in a unit cell is called as the equation of state.
Several different parametrizations of this equation are available. Fitting the energy-volume data to these forms allows us to estimate the equilibrium volume (and in turn, the lattice constant), the bulk modulus, and other related quantities.
One of the most commonly used parametrizations, the Birch-Murnaghan~\cite{birch_finite_1947} equation of state is defined as
\begin{multline}
    E(V) = E_0 + \frac{9B_0 V_0}{16} \left \{
        \left[
            \left(\frac{V_0}{V}\right)^{2/3} - 1
        \right]^{3} B_0'
    \right .
    \\
    \left .
        + \left[
            \left(\frac{V_0}{V}\right)^{2/3} - 1
        \right]^2 \left[
            6 - 4 \left(\frac{V_0}{V}\right)^{2/3}
        \right]
    \right \},
    \label{eq:birch-murnaghan}
\end{multline}
where $E_0$, $V_0$ and $B_0$ denote the equilibrium energy, volume and bulk modulus, respectively, while $B_0'$ denotes the derivative of the bulk modulus with respect to the cell volume $V$, taken at $V_0$.
The Murnaghan~\cite{murnaghan_compressibility_1944} and Vinet~\cite{vinet_temperature_1987} equations are other notable parametric forms.
For the purpose of this paper, we fit the energy-volume data to \cref{eq:birch-murnaghan} using the least-squares optimization.

\section{\label{sec:comp-details} Computational details}
All of our calculations are performed with Gaussian type orbitals (GTO) or their periodic generalizations.~\cite{dovesi_periodic_2000,sun_recent_2020}
For initializing all the calculations, we use PySCF~\cite{sun_pyscf_2018,sun_recent_2020} where we generate the one- and two-electron integrals, and perform mean-field calculations such as Hartree-Fock (HF) or DFT.
For the density-fitting in two-electron integrals, we employ even-tempered Gaussian orbitals~\cite{stoychev_automatic_2017} with progression factor 2.0 for both molecules and extended systems.
For the relativistic calculations, we use \texttt{x2c-$n$ZVPall} ($n=\text{S, T, Q}$) family of basis sets.~\cite{pollak_segmented_2017} In contrast to the traditionally used uncontracted bases for the relativistic calculations, the \texttt{x2c} family of basis sets contains contracted bases minimizing the number of basis functions without a significant sacrifice of overall accuracy. 
To construct the initial Dirac Hamiltonian, the basis set is completely uncontracted (as is also programmed, by default, in PySCF).

\begin{figure*}[tb]
    \centering
    \includegraphics[width=0.9\linewidth]{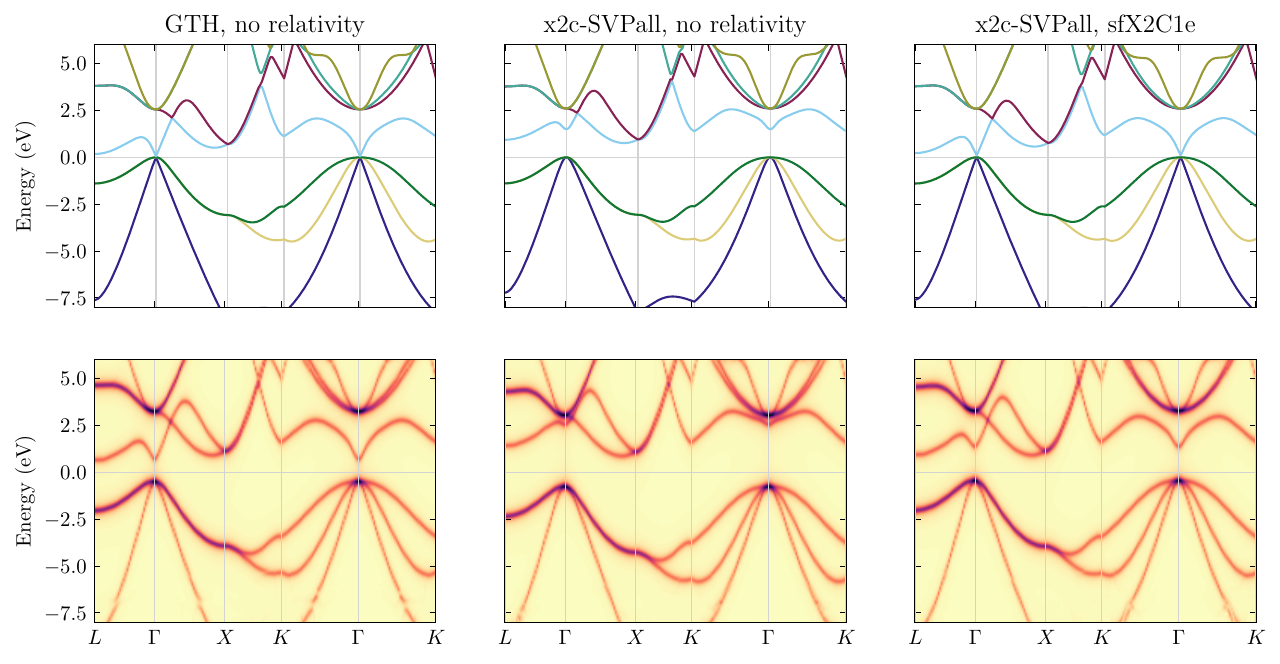}
    \caption{Band structure of germanium calculated with {\em Top panels:} PBE and {\em Lower panels:} \scgw theories, using {\em Left:} \texttt{GTH-PBE} pseudopotential with \texttt{GTH-DZVP-MOLOPT-SR} basis set, {\em Middle:} \texttt{x2c-SVPall} all-electron basis set with no relativistic corrections, and {\em Right:} \texttt{x2c-SVPall} basis set with sfX2C1e Hamiltonian. For the GTH result, relativity is intrinsically accounted for in the pseudopotential. Diamond lattice with lattice constant $a = 5.657 \mathrm{\AA}$ was used, and all calculations were performed with a $6\times 6\times 6$ $k$-mesh sampling in the BZ.}
    \label{fig:germanium-bands}
\end{figure*}

All the dynamical quantities, i.e., imaginary-time and frequency dependent objects, in \scgw are stored using sparse grids.~\cite{grid_orthogonal,grid_spline_cubic,grid_chebychev,grid_super_exponential,li_sparse_2020,shinaoka_efficient_2022}
Sufficiently large grids are used to avoid loss of information in Fourier transforms performed over successive iterations of the \gw cycle.
To obtain spectral functions, IPs and band gaps, the Matsubara Green's function in our calculations is analytically continued to the real-frequency axis.
We use the Nevanlinna analytical continuation to accomplish this task.~\cite{fei_analytical_2021}
To avoid poles on the real axis, a small broadening, a positive imaginary term, is added to the real frequency grid, i.e.,
\begin{equation}
    G (i \omega) \xrightarrow[\text{continuation}]{\text{Nevanlinna analytic}} G (\omega + i \eta).
\end{equation}
For IP and band gap calculations, we employ $\eta = 0.01$ a.u. and $\eta = 0.001$ a.u., respectively. For photo-electron spectra of HgCl$_2$, we have adjusted the chosen a broadening of $\eta = 0.002$ a.u. to match experimental peak widths, while in \cref{fig:germanium-bands}, $\eta = 0.005$ a.u. is used.
All our calculations are converged with rigorous thresholds.
In particular, we ensure that the particle number is converged to $10^{-8}$ and the total energy to at least $10^{-5}$ a.u.
The finite-temperature \scgw code used here is accessible via GitHub.~\cite{iskakov_greenweakcoupling_2023}

\subsection{Need for relativistic corrections}

Before we begin discussing the challenges associated with the application of relativistic \scgw, it is important to justify the necessity of incorporating relativistic effects into correlated all-electron (AE) calculations.
We accomplish this by studying the band structure of germanium.
In \cref{fig:germanium-bands}, we plot three different band structure results evaluated using both PBE and \scgw for germanium with diamond lattice, and lattice constant $a = 5.657 \mathrm{\AA}$. For all these calculations, we used a $6\times 6\times 6$ $k$-mesh sampling in the BZ.

In the top left panel, we employ the GTH basis set and PP, where the relativistic effects are built directly into PP.
Here, we find that PBE predicts a semiconducting phase with 0 eV band gap at the $\Gamma$ point.
The top middle and right panels show PBE calculation with and without the inclusion of scalar relativistic effects, respectively.  Both were performed using an AE basis, namely \texttt{x2c-SVPall}.
Right away, one finds that non-relativistic AE-PBE yields a qualitatively incorrect result with an indirect and sizeable band gap of 1 eV between the $\Gamma$ and $L$ high-symmetry points.
By including scalar relativistic effects at the sfX2C1e level, the band gap is closed resulting in a correct DFT band structure, similar to the GTH results.
For germanium, SOC effects are known to be negligible and therefore, not considered here.

While both top left and top right panels of \cref{fig:germanium-bands} display correct band structure of germanium at the PBE level, the experimental results yield band structure with an indirect band gap of 0.7 eV between $\Gamma$ and $L$ points, which only fortuitously somewhat resembles the band structure from the middle panel evaluated with AE basis and non-relativistic PBE. Consequently, to recover the experimental result for the right reason, both the relativistic treatment and inclusion of correlation beyond the DFT level is necessary. 
This is confirmed by the \scgw results, shown in the lower panels of \cref{fig:germanium-bands}.
In \scgw, the band structure for AE basis set without any relativistic effects is further distorted such that the conduction band, specially near the $\Gamma$-point, is pushed upwards and an indirect band gap larger than 2 eV is predicted between $\Gamma$ and $L$ points.
With appropriate inclusion of relativity, the AE \scgw results provide a better quality band structure, with an indirect band gap of $\sim$1.3 eV, which agrees far better with the experimental value of 0.7 eV.

\section{\label{sec:results} Results}
We are now well-equipped to explore the challenges with relativistic \gw calculations.
In each of the subsections, we describe a specific issue with supporting data.
Further analysis and discussion on these results is presented in the \cref{sec:discussion_conclusion}

\begin{table*}[t]
    \centering
    
    \caption{Comparison of the predicted lattice constants and bulk moduli from calculations with AE basis sets against those with PP. The data highlights the deficiency of PP in describing energy-derivable bulk properties as atomic number $Z$ increases resulting in an increase of the PP approximated core.}
    
    \settowidth{\templen}{Lattice Constant abc xyz 123 456}
    \setlength{\templen}{\dimexpr(\templen-2\tabcolsep)/3}
    
    \begin{tabular}{l|l| *{3}{>{\centering\arraybackslash}p{\templen}} | *{3}{>{\centering\arraybackslash}p{\templen}}}
        \hline
        \multirow{2}{*}{Material} & \multirow{2}{*}{Basis set} & \multicolumn{3}{c|}{Lattice constant ($\mathrm{\AA}$)} & \multicolumn{3}{c}{Bulk modulus (GPa)} \\
        \cline{3-8}
        & & PBE & \scgw & Exp. & PBE & \scgw & Exp. \\
        \hline
        \multirow{3}{*}{Si}
        & \texttt{GTH-DZVP-MOLOPT-SR}           & 5.499 & 5.456 & \multirow{3}{*}{5.431\textsuperscript{\cite{becker_lattice_1982}}}  & 84.8 & 96.6 & \multirow{3}{*}{97.9\textsuperscript{\cite{mcskimin_elastic_1964}}} \\
        & \texttt{x2c-TZVPall} (sfX2C1e)        & 5.494 & 5.413 &  & 83.4 & 102.5 & \\
        & \texttt{def2-TZVP} (non-relativistic) & 5.491 & 5.411 &  & 84.0 & 100.9 & \\
        \hline
        \multirow{2}{*}{Ge}
        & \texttt{GTH-DZVP-MOLOPT-SR}       & 5.812 & 5.733 & \multirow{2}{*}{5.657\textsuperscript{\cite{brummer_anwendung_1972}}} & 52.8 & 66.4 & \multirow{2}{*}{75.0\textsuperscript{\cite{mcskimin_elastic_1963}}} \\
        & \texttt{x2c-TZVPall} (sfX2C1e)    & 5.769 & 5.609 &  & 56.9 & 83.9 & \\
        \hline
        \multirow{2}{*}{$\alpha$Sn}
        & \texttt{GTH-DZVP-MOLOPT-SR}       & 6.711 & 6.649 & \multirow{2}{*}{6.489\textsuperscript{\cite{thewlis_thermal_1954}}} & 52.8 & 32.6 & \multirow{2}{*}{53.1\textsuperscript{\cite{buchenauer_raman_1971}}} \\
        & \texttt{x2c-TZVPall} (sfX2C1e)    & 6.663 & 6.391 &  & 35.2 & 54.3 & \\
        \hline
    \end{tabular}
    
    \label{tab:pseudo-vs-ae}
\end{table*}

\subsection{Deficiency of pseudopotentials}
Modern PPs are designed to account for relativistic effects and, in doing so, eliminate the need for including relativity in subsequent calculations.
The \texttt{GTH} PPs and associated basis sets by Goedecker-Teter-Hutter~\cite{goedecker_separable_1996,hartwigsen_relativistic_1998} constitute one such example.
While the use of PPs generally provide a good description of valence shell band structures, as can be observed from \cref{fig:germanium-bands}, they are far from adequate for other properties, generally derivable from total energy.
We investigate the performance of AE and PP-based basis sets for three materials with diamond-lattice: Si, Ge, and $\alpha$Sn. 
With an increasing atomic number $Z$ and  core size, they serve as excellent representative examples to validate the importance of a proper description of the core orbitals.

In \cref{tab:pseudo-vs-ae}, by comparing PBE and \scgw predictions for lattice constants and bulk moduli against their respective experimental values, we demonstrate that as the atomic number $Z$ of compounds requiring relativistic treatment increases, and consequently the size of their core (approximated by PP) increases, the accuracy of PP-based results decreases, necessitating the use of AE basis sets to achieve experimental agreement.

For Si, which does not require relativistic corrections using PP, the AE basis set has virtually no impact on the quality of results.
On the other hand, for $\alpha$Sn, AE results provide a significant improvement, which is even more pronounced in \scgw than in PBE.
We use \texttt{GTH-DZVP-MOLOPT-SR} basis set with \texttt{GTH-PBE} PP for the PP-based calculations, and \texttt{x2c-TZVPall} basis set, which is the largest basis where we could reliably perform equation-of-state calculations, for AE calculations.

\begin{figure}
    \centering
    \includegraphics[width=0.95\linewidth]{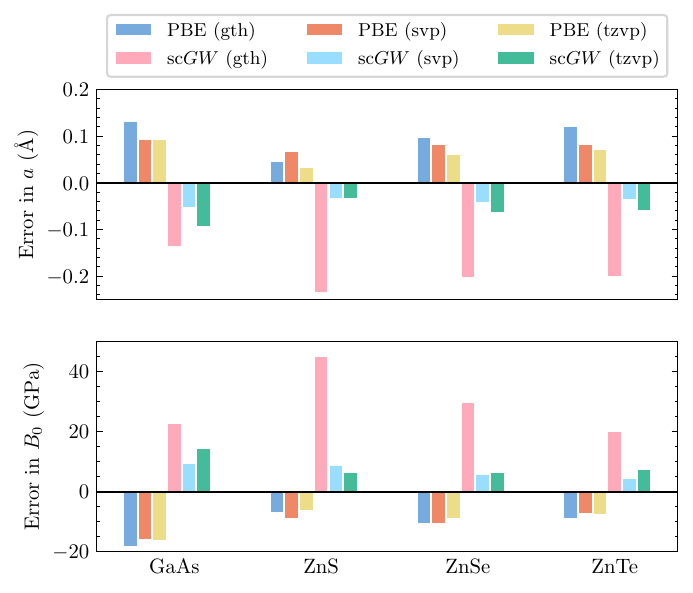}
    \caption{Errors in {\em Top:} lattice constants $a$, and {\em Bottom:} bulk moduli $B_0$ for selected materials. All compounds are calculated with \texttt{GTH-DZVP-MOLOPT-SR}, \texttt{x2c-SVPall} and \texttt{x2c-TZVPall} basis sets. Both AE-PBE and AE-\scgw results are reported using the sfX2C1e Hamiltonian, with a $4\times 4\times 4$ $k$-mesh sampling. Experimental values for lattice constants and bulk moduli can be found in Refs.~\cite{fukumori_measurements_1988,nelmes_chapter_1998} and \cite{cline_volume_1965,blakemore_semiconducting_1982,pellicer-porres_high-pressure_2005}, respectively.}
    \label{fig:latts_and_bulk_results}
\end{figure}

For Si and other elements where the size of atomic core is small and the relativistic effects are negligible PPs are usually capable of yielding very good results.  
However, as the atomic number $Z$ increases the problems with PPs become more noticeable as illustrated  for Germanium and subsequently  $\alpha$-Sn.
In \cref{fig:latts_and_bulk_results}, we showcase more examples where \scgw evaluated in AE basis sets provides highly accurate results for both equilibrium lattice constants and bulk moduli. In contrast, the \scgw results evaluated using GTH PPs are significantly worse than those obtained with PBE based on GTH PP. 
Particularly, in ZnX (X=S, Se, Te), AE relativistic \scgw is highly effective in contrast to the GTH results.
This poor performance of GTH PPs with \scgw can be attributed to missing correlation contribution arising from the overlap of zinc's inner $p$ and $d$ orbitals with the inner $p$ orbitals of the chalcogen.

In Green's function calculations such as \scgw, when AE basis sets are employed, the core orbitals are described by self-energy matrix elements that have both static and dynamic parts. In contrast, when PPs are used, the dynamic part of self-energy is entirely missing from the description of core orbitals.
For light elements such as Si, the magnitude of the dynamic self-energy for the core orbitals is small, and in addition, there are also only a few core orbitals. Consequently, the error of neglecting these self-energies is overall insignificant.
On the other hand, given the large number of core orbitals in heavy elements, the number of dynamic core-core and core-valence self-energy matrix elements neglected by PPs is significant.
Moreover, the mixing between valence and outer-core orbitals becomes important, resulting in sizable values of the outer core and core-valence self-energy matrix elements.
Since the calculations involving PPs, by design, cannot include any dynamic self-energy for the core, it is reasonable to expect that for elements where such self-energy contributions are significant, the use of PPs will result in sizeable errors.
These errors can be additionally compounded by a lack of good, well optimized PPs for heavier elements. 
Therefore, it is worth noting that for Green's function methods, the use of PPs poses a different set of problems than for DFT. The DFT results can only be affected by a lack of well optimized PPs, while Green's function methods would always be penalized by the absence of dynamic self-energy even for the best optimized PPs. For some lighter elements, one can only hope that these dynamic self-energy matrix elements have insignificant values, however, this assumption cannot be generally fulfilled.

\subsection{Convergence issues in basis sets}
For heavy elements, the very large size of the one-electron basis sets effectively hinders the ability to investigate convergence with respect to basis set size in AE calculations.
While for molecules, this formidable challenge still can be overcome, it becomes insurmountable in calculations of solids.
The problem is relatively mild, often almost non-existent, at the level of DFT or HF, but quickly becomes aggravated for correlated calculations such as \scgw.
This can also be understood by the fact that basis sets are usually optimized at the atomic level using SCF calculations.
Such behavior has also been observed by Kutepov in Ref.~\onlinecite{kutepov_ground_2017}, where DFT calculations converged far more rapidly than \scgw while using linearized augmented-plane-wave (LAPW) basis functions.
Here, we examine such convergence issues in detail.

\begin{table}[tb]
    \caption{PBE0 and \scgw results for ionization potentials (in eV) for selected silver halids. The results were evaluated using different basis sets and hten extrapolated to the basis set limit.  Notice that the mean-field method appears to be converged already at the triple-$\zeta$ level.}
    \label{tab:mol_ip_convergence}
    \centering
    
    \begin{tabular}{l|cc|cc|cc}
        \hline
        \multirow{2}{*}{basis}  & \multicolumn{2}{c|}{AgCl} & \multicolumn{2}{c|}{AgBr} & \multicolumn{2}{c}{AgI}   \\ \cline{2-7}
                                & PBE0       & \scgw        & PBE0       & \scgw        & PBE0       & \scgw        \\ \hline
        \texttt{x2c-TZVPall}    & 7.51       & 9.54         & 7.18       & 9.11         &  6.77      & 8.46         \\
        \texttt{x2c-QZVPall}    & 7.58       & 9.78         & 7.23       & 9.29         & 6.83       & 8.64         \\
        Extrapolated            & 7.67       & 10.09        & 7.29       & 9.53         & 6.89       & 8.86         \\ \hline
    \end{tabular}
\end{table}

\begin{figure}
    \centering
    \includegraphics[width=0.95\linewidth]{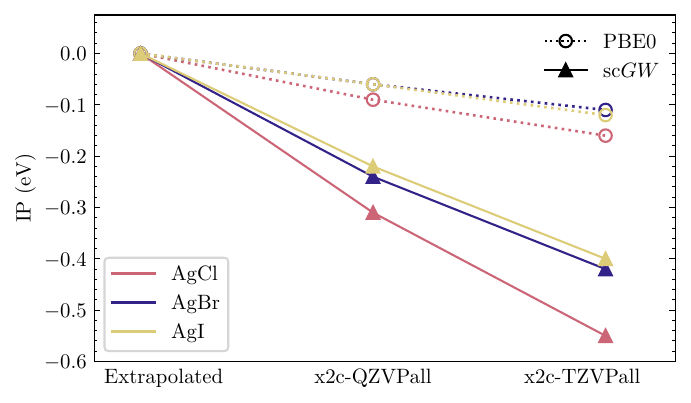}
    \caption{Graphical representation of convergence trends in \cref{tab:mol_ip_convergence}. All IPs are reported relative to the respective extrapolated values, which shows that \scgw exhibits a much slower convergence than PBE0.}
    \label{fig:ip_convergence}
\end{figure}

First, we look at IPs for AgCl, AgBr and AgI, which are representative examples of closed-shell relativistic molecules with an increasing SOC as we go from Cl to I.
These systems have been widely used to benchmark recent implementations of relativistic \gw methods.~\cite{scherpelz_implementation_2016,forster_two-component_2023,abraham_relativistic_2024}
For these molecules, in \cref{tab:mol_ip_convergence}, we show IPs evaluated in PBE0 and \scgw with \texttt{x2c-TZVPall} and \texttt{x2c-QZVPall} basis sets.
Using these two IP values, an extrapolated IP value (also reported in ~\cref{tab:mol_ip_convergence}) is calculated assuming an inverse relationship between IP and the number of AOs in the respective basis set.
We use the scalar relativistic Hamiltonian in all these calculations.
Comparing these trends graphically in \cref{fig:ip_convergence}, we can clearly observe that for PBE0, the extrapolated IPs are within 0.1 eV of \texttt{QZ}, displaying a fast convergence.
However, for \scgw, the quadruple-$\zeta$ results are far from converged, with extrapolation leading to changes of more than 0.2 eV.
It is worth mentioning that an extrapolation based on two calculations, triple- and quadruple-$\zeta$, is already far from ideal, specially when the resultant corrections are so large.
Having quintuple-$\zeta$ results would greatly enhance the quality of results.
However, even not considering the computational challenge that such a calculation poses, in many relativistic basis--sets, including x2c, quintuple-$\zeta$ bases are simply not available.

\begin{table*}[bhtp]
    \begin{tabular}{l|c|cc|cc|cc|cc}
        \hline
        \multirow{2}{*}{basis} & \multirow{2}{*}{orbitals} & \multicolumn{2}{c|}{$L-L$} & \multicolumn{2}{c|}{$\Gamma-\Gamma$} & \multicolumn{2}{c|}{$X-X$} & \multicolumn{2}{c}{$L-\Gamma$} \\
        \cline{3-10}
        & & PBE & \scgw & PBE & \scgw & PBE & \scgw & PBE & \scgw \\
        \hline
        \textbf{AgBr}   &       &       &       &           &       &           &       &       &       \\
        x2c-SVPall      & 85    & 3.887 & 6.73  & 2.536     & 5.38  & 3.926     & 7.68  & 0.699 & 3.71  \\
        x2c-TZVPall     & 111   & 3.833 & 6.50  & 2.529     & 5.23  & 3.901     & 7.49  & 0.688 & 3.54  \\
        x2c-QZVPall     & 167   & 3.776 & 6.31  & 2.546     & 5.00  & 3.886     & 7.17  & 0.678 & 3.34  \\
        \hline
        \textbf{CdSe}   &       &       &       &           &       &           &       &       &       \\
        x2c-SVPall      & 88    & 3.209 & 5.53  & 0.663     & 2.73  & 5.006     & 6.98  & 1.428 & 3.60  \\
        x2c-TZVPall     & 111   & 3.193 & 5.46  & 0.659     & 2.69  & 5.004     & 6.94  & 1.433 & 3.55  \\
        x2c-QZVPall     & 192   & 3.179 & 5.30  & 0.633     & 2.56  & 5.003     & 6.87  & 1.406 & 3.40  \\
        \hline
    \end{tabular}
    \caption{Convergence of PBE and \scgw band gaps for AgBr and CdSe, with respect to basis sets from the \texttt{x2c} family. The results were calculated with the sfX2C1e-Coulomb Hamiltonian and a $4\times4\times4$ $k$-mesh and inverse temperature $\beta=300$ a.u.$^{-1}$ (for \scgw). The total numbers of GTOs per cell for each basis set are listed in the second column.
    \label{tab:bandgap_basis_convergence}}
\end{table*}

Next, we look at the band structure in relativistic solids, with focus on AgBr and CdSe.
For AgBr, rock salt crystal structure with $a = 5.774 \mathrm{\AA}$ is considered, while the zincblende phase in CdSe with $a = 6.05 \mathrm{\AA}$ is used.
All calculations were performed using a $4\times 4 \times 4$ $k$-mesh sampling, using \texttt{x2c-SVPall}, \texttt{x2c-TZVPall} and \texttt{x2c-QZVPall} basis sets.
An inverse temperature of $\beta = 300$ a.u.$^{-1}$ was used.
For the \texttt{QZVPall} basis in AgBr, diffuse $s$ and $p$ orbitals with Gaussian exponents smaller than 0.05 were removed to avoid linear dependencies in the basis set.
High angular momentum $g$-orbitals were removed from \texttt{QZVPall} basis set in both systems.

For AgBr, in Ref.~~\onlinecite{yeh_relativistic_2022}, Yeh et al. noted poor convergence of band gap values along the path containing special symmetry points. 
Here we present further data contrasting \scgw convergence trends against PBE.

In \cref{tab:bandgap_basis_convergence}, we present direct and indirect band gaps in AgBr and CdSe, with \texttt{x2c-SVPall}, \texttt{x2c-TZVPall} and \texttt{x2c-QZVPall} basis sets.
Already at the triple-$\zeta$ level, for both systems, PBE band gaps are well converged to within 0.05 eV when compared with quadruple-$\zeta$ basis.
However, \scgw once again shows much slower convergence trends, and going from triple- to quadruple-$\zeta$ still introduces corrections larger than 0.1 eV for CdSe and 0.2 eV for AgBr.

Finally, in \cref{tab:latt_convergence} we show convergence trends for lattice constants for Si, Ge, GaAs, $\alpha$Sn, and InSb.
All these systems have a diamond (zincblende for GaAs and InSb) structure, with an increasing size of atomic core as we go from Si to InSb.
Similar to previous observations, the convergence of \scgw values is more than 2-3 times slower than for PBE.
At the level of DFT, it seems that only for Si the larger basis set marginally improves  the agreement with experiment, while for all the other materials, the results either do not change or become worse.
Consequently, one can suspect that DFT optimization of higher level bases is not as effective as lower level bases since the convergence with respect to the basis size is achieved relatively early on.
This suspicion is somewhat confirmed when examining the sc$GW$ results. Almost all \texttt{SVPall} sc$GW$ results are surprisingly close to experiment and do not seem to be improved in a larger basis.
For periodic systems, an extrapolation procedure, similar to the one used for molecular IPs, cannot be considered. This is because SVP is not a sufficiently large basis and using it together with only the TZVP basis would lead to largely inaccurate extrapolated data. 
While for molecular systems, QZ-level calculations remain accessible and make extrapolations possible, for the periodic systems, these calculations are extremely expensive and mostly out of reach.

\begin{table}[tb]
    \centering
    \caption{Basis set convergence trends in lattice constants [\AA] for selected compounds. Results for \texttt{x2c-SVPall} and \texttt{x2c-TZVPall} basis sets are shown, denoted by SVP and TZVP, respectively. For readers' convenience, we also list the difference between them.}
    \label{tab:latt_convergence}
    
    \begin{tabular}{l|ccc|ccc|c}
        \hline
        \multirow{2}{*}{System} & \multicolumn{3}{c|}{PBE}              & \multicolumn{3}{c|}{\scgw}        & \multirow{2}{*}{Exp.}                     \\ \cline{2-7}
                                & SVP       & TZVP          & Diff.     & SVP       & TZVP      & Diff.     &                                           \\ \hline
        Si                      & 5.504	    & 5.494         & -0.010    & 5.453     & 5.413     & -0.04     & 5.431\textsuperscript{\cite{becker_lattice_1982}}         \\
        Ge                      & 5.779     & 5.773         & -0.006    & 5.622     & 5.559     & -0.063    & 5.657\textsuperscript{\cite{brummer_anwendung_1972}}      \\
        GaAs                    & 5.747     & 5.748         & 0.001     & 5.600     & 5.559     & -0.041    & 5.653\textsuperscript{\cite{fukumori_measurements_1988}}  \\
        $\alpha$-Sn             & 6.620      & 6.663         & 0.043    & 6.456     & 6.391     & -0.065    & 6.489\textsuperscript{\cite{thewlis_thermal_1954}}        \\
        InSb                    & 6.597     & 6.64          & 0.043     & 6.463     & 6.381     & -0.082    & 6.479\textsuperscript{\cite{straumanis_lattice_1965}}     \\
        \hline
    \end{tabular}
\end{table}

\begin{figure*}[htb]
    \centering
    \begin{minipage}[t]{0.4\linewidth}
        \settowidth{\templen}{$B_0$ (GPa)}
        \vspace{7pt}
        \begin{tabular}[b]{l *{2}{|>{\centering\arraybackslash}p{\templen}}}
            \hline
            Method                          &   a ($\mathrm{\AA}$)      &   $B_0$ (GPa)         \\
            \hline 
            PBE                             &   5.491                   &   84.01               \\
            PBE w/o diffuse orbs.           &   5.378                   &   124.04              \\
            \scgw                           &   5.411                   &   100.93              \\
            \scgw w/o diffuse orbs.         &   5.306                   &   146.88              \\
            \hline
            Experiment\textsuperscript{\cite{becker_lattice_1982,mcskimin_elastic_1964}} &   5.431   &   97.8    \\
            \hline
        \end{tabular}
    \end{minipage}
    \begin{minipage}[t]{0.5\linewidth}
        \vspace{0pt}
        \includegraphics[width=0.9\linewidth]{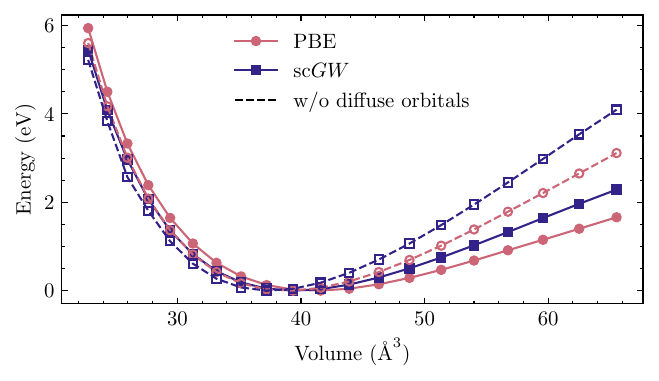}
    \end{minipage}
    \caption{Left panel: Lattice constants $a$ and bulk modulus $B_0$ for silicon, calculated using \texttt{def2-TZVP} basis with and without diffuse orbitals (Gaussian exponent $< 0.1$). Both PBE and \scgw results, along with experimental values, are shown. Non-relativistic Hamiltonian was employed for these calculations. Right panel: The corresponding energy-volume curves. }
    \label{fig:eos_diffuse}
\end{figure*}

\subsection{Linear dependencies}
For correlated calculations, the quest for converging properties to a complete basis set limit is not only affected by slower convergence and lack of larger basis sets but also by linear dependencies arising in the large basis sets.
Particularly in solids, due to close packing of atoms, adding more basis functions introduces strong linear dependencies among the atomic orbitals.
As a result, the condition number of the overlap matrix becomes too large even for an easy execution of mean-field methods.
The problem of linear dependencies is not unique to GTOs considered here, and has been reported in other kinds of orbital representations, namely Slater-type orbitals (STO), and also in LAPW.~\cite{kotani_fusion_2010,jiao_application_2015,karsai_importance_2017,forster_low-order_2021}

Even though linear dependencies are more severe in ionic compounds, they also remain problematic in covalent compounds (such as silicon) and arise at small bond distances in all compounds.
One way to remedy this problem in GTO basis sets is to simply remove the diffuse functions.
These orbitals have long tails because of their small Gaussian exponents, leading to an increased overlap with other atomic orbitals.
However, when studying bulk properties, such a procedure generally does more harm than good.
When the bonds are stretched, the lack of diffuse orbitals leads to a deficient description of atomic bonds and substantially alters the equation of state, causing significant errors in both the equilibrium lattice constant and bulk modulus.
This phenomenon is investigated for silicon in \cref{fig:eos_diffuse}, where we have considered a rather simple case of non-relativistic calculation involving the widely used \texttt{def2-TZVP} basis set.
Here, all orbitals with Gaussian exponents smaller than 0.1 were dropped, leading to lattice constants differing by more than 0.1 $\mathrm{\AA}$ and over 50 \% error in bulk modulus.
Note that an ad-hoc re-introduction of diffuse functions as we stretch the bonds introduces jumps in the energy-volume curve, precluding a reliable fit to the Birch-Murnaghan equation.
Even though the computational cost is still affordable, due to arising linear dependencies, converging DFT and \gw calculations for Si with a larger basis set proves extremely challenging and is therefore not pursued here.

\subsection{Lack of accurate experimental reference}
\begin{figure}[tb]
    \centering
    \includegraphics[width=0.95\linewidth]{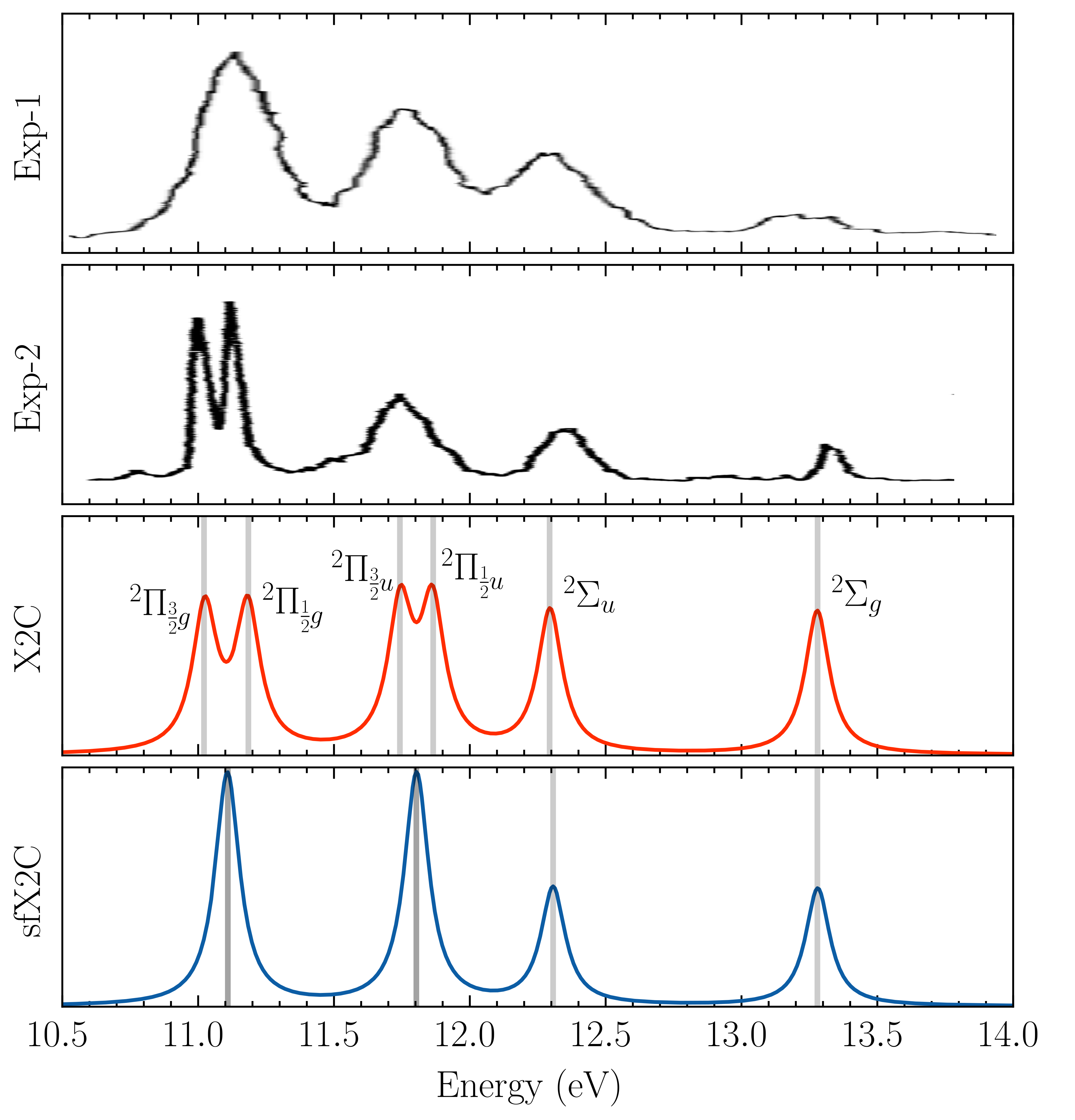}
    \caption{Comparison of experimental spectra with \scgw results for HgCl$_2$, obtained with both scalar and two-component relativistic effects. For these calculations, we have used the \texttt{x2c-QZVPall} basis. Both experimental spectra, exp-1 by Bogges et al.~\cite{bogges_photoelectron_1973} and exp-2 by Eland et al.~\cite{eland_photoelectron_1970}, are shifted by 0.35 eV to match the associated IP peaks. }
    \label{fig:hgcl}
\end{figure}

To benchmark new theoretical methods that are pushing the limits of accuracy and applicability, reliable experimental references are necessary.
Here, we consider the HgCl$_2$ molecule as an example to highlight a potential area were better experimental data can provide feedback to improve theoretical methods.
In \cref{fig:hgcl}, we compare \scgw results with two different experimental spectra for HgCl$_2$.
In one of the experiments by Borgges et al.~\cite{bogges_photoelectron_1973} (labeled Exp-1 in the plot), the SOC splitting is unresolved in both first and second IPs.
Results by Eland et al.~\cite{eland_photoelectron_1970} (Exp-2 in \cref{fig:hgcl}) clearly resolve the SOC splitting in the first IP (${}^2\Pi_g$ state), but reports difficulty with the second peak (${}^2\Pi_u$ state).
Among $GW$ results, scalar relativistic corrections (sfX2C1e) capture the overall peak structure, producing a spectra similar to Exp-1.
However, the spin-free theory understandably misses out on the SOC splitting which, in turn, is captured at the X2C1e level.
In fact, for \scgw with X2C1e, spin-orbit splitting is visible for both ${}^2\Pi_g$ and ${}^2\Pi_u$ states.
For the first ionization peak, \scgw predicts a 0.16 eV splitting for ${}^2\Pi_{\frac{3}{2}g}$, ${}^2\Pi_{\frac{1}{2}g}$ states which is close to the experimental value of 0.12 eV.
Similarly, for the ${}^2\Pi_{\frac{3}{2}u}$, ${}^2\Pi_{\frac{1}{2}u}$ state, a splitting of 0.12 eV is predicted, but the corresponding results are not resolved in experiments.

\section{\label{sec:discussion_conclusion} Discussion and conclusions}
We have presented ample numerical evidence that when employing Green's function methods, particularly \scgw, the current frameworks of PPs and orbital representation encounter challenges in fully realizing the potential of precise quantum chemistry calculations for relativistic systems.
One of the challenges is that PPs are not adept in accurately describing properties other than valence-shell electronic structure such as band gaps and ionization potentials.
When a material is pressurized, e.g., while studying the equation of state in a material, the correlation effects arising from the overlap between atomic cores cannot be captured by PPs, and highly expensive AE calculations become unavoidable.
Effective cores and PPs where a smaller number of electrons are modelled as part of the core offer one way to overcome this hurdle, and while some candidates are available,~\cite{roy_revised_2008,burkatzki_energy-consistent_2008,weigand_relativistic_2014,zaitsevskii_generalized_2023,hill_correlation_2022} more developments and systematic benchmarks of new ECPs and PPs are desirable.

Additionally, it should be noted that when using PPs in Green's function methods, there is an inherent error that comes from the absence of dynamic self-energy matrix elements for core and core-valence sectors even if the best optimized PPs are employed. Especially for heavier elements, one can surmise that such a PP approximation will lead to significant errors since both the values and the number of outer core dynamic self-energy matrix elements may be large. Consequently, even with the most optimized PPs, as they are formulated now, we can expect significant errors. 

For correlated methods such as \scgw understanding the basis set convergence is a bit difficult.
Even when well designed and optimized basis sets are available,~\cite{dunning_gaussian_1989}, correlated methods generally exhibit a slower convergence than mean-field ones.
Yet, the fact that basis sets are optimized at the level of mean-field cannot be overlooked and, in fact, is potentially one the reasons behind faster convergence in DFT and HF.~\cite{heyd_energy_2005,lee_approaching_2021}
For lighter elements, basis set convergence is easily achieved and such problems do not require any particular attention.
However, same is not true for heavy elements where relativity is also important.
The issues with basis sets for heavy elements is well known and has been a topic of recent research.~\cite{rueda_espinosa_novel_2023,Zhang2022}

Furthermore, most quantum chemistry basis sets have been optimized for molecular applications.
When these are applied to solids, not only the molecular level optimization may prove insufficient, additional problems related to linear dependency and unfavorable convergence trends are more prone to arise.
Recently, Ye et al.~\cite{ye_correlation-consistent_2022} also noticed similar problems and proposed correlation-consistent basis sets optimized for solids.
Another way to improve the quality of basis sets is to perform a material-specific optimization.~\cite{zhou_material-specific_2021}
These developments, including other means to eliminate linear dependencies,~\cite{lehtola_curing_2019,lehtola_accurate_2020} are yet to be applied to relativistic materials and molecules.
It is also likely that a better one-particle orbital representation, with strictly orthogonal atomic orbitals, may be required to simultaneously overcome the issues related to convergence and linear-dependencies.~\cite{white_multisliced_2019,qiu_hybrid_2021,white_nested_2023}
Consequently, one can expect that in the future explicitly orthogonal single-particle orbitals, such as tensor-train numerical STOs,~\cite{jolly_tensorized_2023} or orthogonal Gausslets,~\cite{white_multisliced_2019,qiu_hybrid_2021,white_nested_2023} will show promise to solve the problem of linear dependency.
Constructing PPs and basis functions from correlated calculations also may offer a viable route to improve the reliability of relativistic calculations.

Lastly, we also show that as theoretical methods become more accurate, better reliable experimental benchmarks data with well resolved photo-electron spectra are desirable to both validate theoretical results and push for new theoretical developments.

\begin{acknowledgments}
    This material is based upon work supported by the U.S. Department of Energy, Office of Science, Office of Advanced Scientific Computing Research and Office of Basic Energy Sciences, Scientific Discovery through Advanced Computing (SciDAC) program under Award Number DE-SC0022198.
    This research used resources of the National Energy Research Scientific Computing Center, a DOE Office of Science User Facility supported by the Office of Science of the U.S. Department of Energy under Contract No. DE-AC02-05CH11231 using NERSC awards BES-ERCAP0029462 and BES-ERCAP0024293.
\end{acknowledgments}

\appendix

\bibliography{paper}

\end{document}